\documentclass[preprint,aps]{revtex4-1}
\pdfoutput=1

\usepackage{amsmath,amssymb,amsfonts,dcolumn,color,graphicx,graphics,latexsym,placeins,epsfig}
\usepackage{epsfig}
\usepackage{bm}
\usepackage{slashed}
\usepackage{latexsym}
\usepackage{natbib}
\usepackage{url}
\usepackage{dcolumn}
\usepackage{color}
\usepackage{amsfonts,amssymb,amsmath}
\usepackage{graphicx,epsfig}
\usepackage{psfrag}
\usepackage{subfigure}
\usepackage{tabularx}
\usepackage{hyperref}
\hypersetup{colorlinks=true}

\newcommand{\be}{\begin{equation}}
\newcommand{\ee}{\end{equation}}
\newcommand{\ba}{\begin{eqnarray}}
\newcommand{\ea}{\end{eqnarray}}

\begin{document}

\title{Constraints on monopole-dipole potential from the tests of gravity }
 \author{Tanmay Kumar Poddar}
\email[Email Address: ]{tanmay.poddar@tifr.res.in}
\affiliation{Department of Theoretical Physics, Tata Institute of Fundamental Research, Mumbai - 400005, India}

\author{Debashis Pachhar}
\email[Email Address: ]{debashispachhar@prl.res.in}
\affiliation{Theoretical Physics Division, 
Physical Research Laboratory, Ahmedabad - 380009, India}
\affiliation{Discipline of Physics, Indian Institute of Technology, Gandhinagar - 382355, India}
\begin{abstract}
Ultralight Axion Like Particle (ALP) can mediate a long range monopole-dipole macroscopic force between Earth and Sun if Earth is treated as a polarized source. There are about $10^{42}$ number of polarized electrons in Earth due to the presence of the geomagnetic field. The monopole-dipole interactions between electrons in Earth and nucleons in Sun can influence the perihelion precession of Earth, gravitational light bending and Shapiro time delay. The contribution of monopole-dipole potential is limited to be no larger than the measurement uncertainty. We obtain the first bound on monopole-dipole strength from single astrophysical observations. The perihelion precession of Earth puts the stronger bound on monopole-dipole coupling strength as $g_Sg_P\lesssim 1.75\times 10^{-16}$ for the ALP of mass $m_a\lesssim 1.35\times 10^{-18}~\rm{eV}$. We also obtain constraints on monopole-dipole coupling strength as $g_Sg_P\lesssim 5.61\times 10^{-38}$ from two different astrophysical observations such as the perihelion precession of the planet and the red giant branch. The bound is three orders of magnitude stronger than the Eot-Wash experiment and one order of magnitude stronger than the $(\rm{Lab})^N_S\times (\rm{Astro})^e_P$ limit. 
\end{abstract}
\pacs{}
\maketitle
\section{Introduction}
Ultralight pseudoscalar bosons such as Axion Like Particles (ALPs) can mediate a long range macroscopic force between two objects if the mass of the ALP is smaller than the inverse distance between the two bodies \cite{Moody:1984ba}. Unlike QCD (Quantum Chromodynamics) axions \cite{Peccei:1977hh,Weinberg:1977ma,Wilczek:1977pj,Peccei:1977ur}, the ALPs are generated by string compactifications \cite{Svrcek:2006yi}. The mass of the ALP and its symmetry breaking scale are independent of each other. These ultralight bosons couple very weakly with the Standard Model (SM) particles and hence, it is extremely challenging to search these particles in direct detection experiments. Nevertheless, there are many ongoing and future experiments which are built to probe these particles. Several laboratories, astrophysical, and cosmological constraints on the mass and decay constant of QCD axions and ALPs are discussed in \cite{Inoue:2008zp,Arik:2008mq,Hannestad:2005df,Melchiorri:2007cd,Hannestad:2008js,Hamann:2009yf,Semertzidis:1990qc,Cameron:1993mr,Robilliard:2007bq,Chou:2007zzc,Sikivie:2007qm,Kim:1986ax,Cheng:1987gp,Rosenberg:2000wb,Hertzberg:2008wr,Visinelli:2009zm,Battye:1994au,Yamaguchi:1998gx,Hagmann:2000ja,KumarPoddar:2019jxe,KumarPoddar:2019ceq,Poddar:2020qft,Poddar:2021sbc,Poddar:2021ose}. Ultralight ALPs can also be a promising candidate for Dark Matter (DM) \cite{Preskill:1982cy,Abbott:1982af,Dine:1982ah}. The mass of the ALP can be as small as $10^{-22}~\rm{eV}$ and the corresponding de Broglie wavelength is of the order of the size of a dwarf galaxy $(1-2~\rm{kpc})$ \cite{Hu:2000ke,Hui:2016ltb}. Therefore, the ultralight particle DM behaves as a wave. Such wave DM can solve the longstanding core-cusp problem \cite{Oh:2010ea,Hlozek:2014lca,Mocz:2018ium} and evade the DM direct detection constraints \cite{XENON:2018voc,LUX:2013afz,PandaX-II:2017hlx}. Hence, phenomenologically it is important to search these particles and obtain constraints on ALP parameters. 

The mediation of ALP between two fermion currents can give rise to long range macroscopic forces. Generally, the ALP can couple with fermions either by a spin dependent pseudoscalar coupling $(\bar{\psi}\gamma_5\psi a)$ or by a spin independent scalar coupling $(\bar{\psi}\psi a)$, where $\psi$ is the fermion field and $a$ is the axion field. The spin independent scalar current-current interaction and spin dependent pseudoscalar current-current interaction with one ALP exchange can give rise to the usual monopole-monopole and dipole-dipole potential respectively. However, if ALP couples with fermion currents by scalar coupling at one vertex and pseudoscalar at another vertex, then the ALP can mediate a long range monopole-dipole potential. The monopole-monopole, and dipole-dipole forces are parity $(P)$ and time reversal $(T)$ conserving. However, the search for monopole-dipole force is interesting as it can violate $P$ and $T$.
\begin{figure}[h]
\includegraphics[height=6cm]{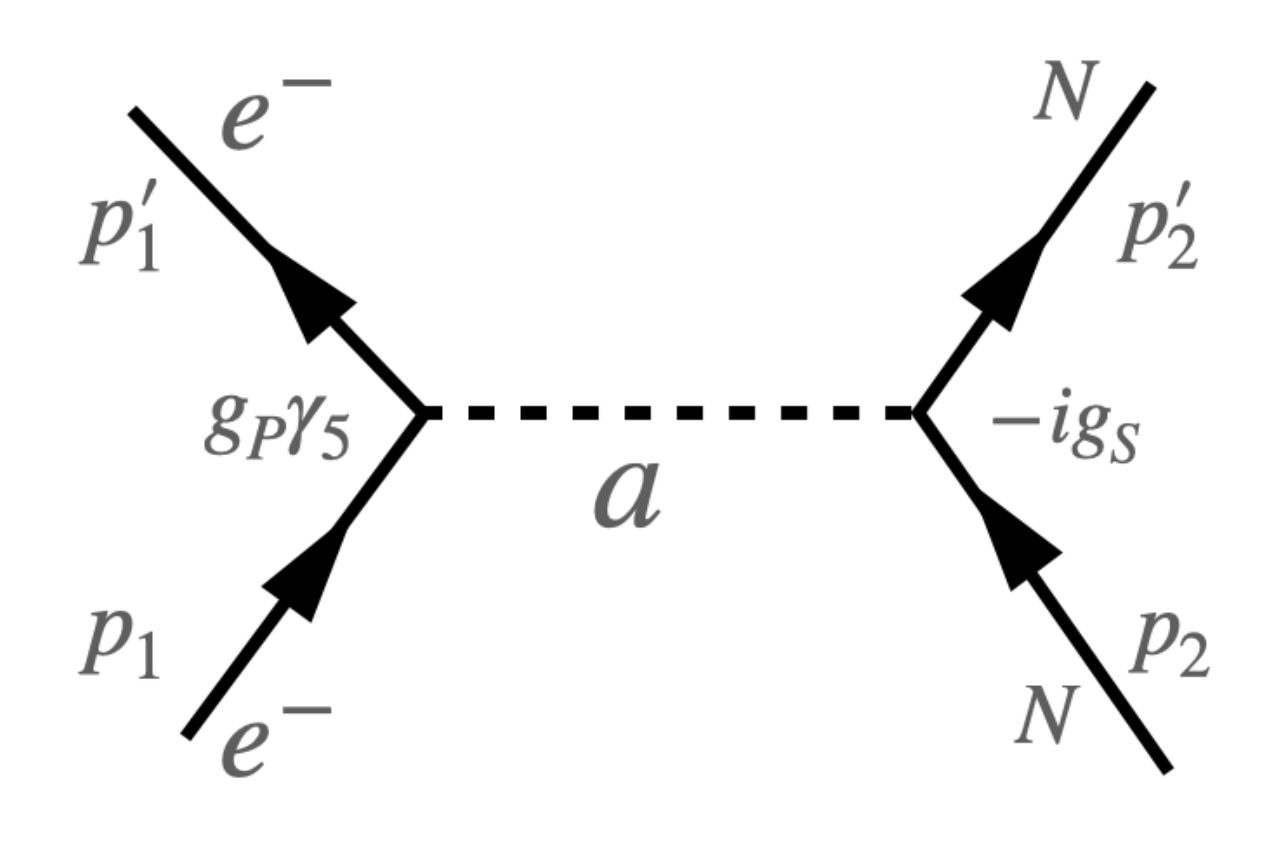}
\caption{Feynman diagram of $e^-N$ scattering mediated by pseudoscalar ALP. Here, the electrons are polarized and the nucleons are unpolarized.}
\label{feynman}
\end{figure} 

The Feynman diagram for an axion mediated monopole-dipole potential between polarized electron current and unpolarized nucleonic current is shown in FIG. \ref{feynman}. The expression of the monopole-dipole potential mediated by ultralight ALP between two fermion currents is given by \cite{Moody:1984ba,Daido:2017hsl}
\begin{equation}
V(r)=\frac{g_Pg_S}{4\pi m_e}(\mathbf{s_1}.\hat{r})\Big(\frac{m_a}{r}+\frac{1}{r^2}\Big)e^{-m_a r},
\label{eq:ak}
\end{equation}
where we consider that ALP with mass $m_a$ is coupled with the polarized electron by a pseudoscalar coupling with strength $g_P$ and the ALP is also coupled with unpolarized nucleon by a scalar coupling with strength $g_S$. Here, $m_e$ is the mass of the electron, and $\mathbf{s_1}$ is the electron's spin vector. The term $\mathbf{s_1}.\hat{r}$ violates $P$ and $T$ symmetries. The derivation of Eq. \ref{eq:ak} is given in Appendix~\ref{app1}.

There are various experiments which are trying to search for parity and time reversal violating monopole-dipole potential \cite{OHare:2020wah,Crescini:2017uxs,Agrawal:2022wjm,Davoudiasl:2022gdg,Stadnik:2017hpa}. Such potential can be constrained from the torsion balance method using polarized electrons in the torsion pendulum and unpolarized nucleons in Earth and Sun \cite{Heckel:2008hw}. The bound on $g_Sg_P$ obtained from this laboratory experiment is most sensitive for the axions of mass $m_a\lesssim 10^{-14}~\rm{eV}$. The QUAX-$g_Sg_P$ \cite{Crescini:2016lwj,Crescini:2017uxs} experiment obtains lab-lab bound on $g_Sg_P$ for the mass of the axion $5\times 10^{-7}~\rm{eV}\lesssim m_a\lesssim 10^{-5}~\rm{eV}$. An experiment like ARIADNE is made to search for monopole-dipole potential using a laser polarized ${}^3He$ and a rotating tungsten source mass \cite{ARIADNE:2017tdd}. This lab-lab $g_Sg_P$ bound is valid for axions of mass $1~\rm{\mu eV}\lesssim m_a\lesssim 6~\rm{meV}$. In \cite{Afach:2014bir}, polarized ultracold neutron spins and unpolarized nucleons are used to constrain such potential. This lab-lab experiment can probe axions of mass $1~\rm{meV}\lesssim m_a\lesssim 0.1~\rm{eV}$. There are other laboratory experiments like SMILE $(m_a\lesssim 10^{-10}~\rm{eV})$ \cite{Lee:2018vaq}, NIST $(m_a\lesssim 10^{-14}~\rm{eV})$ \cite{Wineland:1991zz}, Washington $(10~\rm{\mu eV}\lesssim m_a\lesssim 10~\rm{meV})$ \cite{Hoedl:2011zz,Terrano:2015sna}, Magnon based axion dark matter search $(m_a\lesssim 10^{-5}~\rm{eV})$ \cite{Chigusa:2020gfs,Mitridate:2020kly} which obtain bounds on monopole-dipole interaction. The cooling of red giants and white dwarfs put constraint on $g_P$ as $g_P\lesssim 1.6\times 10^{-13}$ \cite{Capozzi:2020cbu} and the constraint on $g_S$ obtained from the energy loss of globular clusters stars is $g_S\lesssim 1.1\times 10^{-12}$ \cite{Hardy:2016kme}. Multiplying these two numbers, one can obtain the bound on monopole-dipole coupling as $g_Sg_P\lesssim 10^{-25}$ for $m_a\lesssim 10~\rm{keV}$. The lab-astro bound on $g_Sg_P$ is obtained from two independent experimental bounds and the bound is sensitive for $m_a\lesssim 10^{-18}~\rm{eV}$. The astro-astro $g_Sg_P$ bound also considers two separate observations.

So far there is no single astrophysical phenomenon that can directly constrain the monopole-dipole interaction. The most stringent bound on monopole-dipole interaction is claimed by combining the best experimental bound on scalar interaction multiplied by the best astrophysical bound from stellar energy loss on the pseudoscalar interaction. It is also highlighted by the authors of \cite{OHare:2020wah} that in several scenarios these hybrid bounds could be overly stringent leading to a premature abandoning of the axions as an attractive theoretical prospect. There is a lack of a complete astrophysical probe of monopole-dipole potential as most of the astrophysical objects are considered to be unpolarized. Even if a polarized astrophysical object is considered, its degree of polarization is not known precisely. 

In this paper, we consider the Earth as a polarized source and there are about $10^{42}$ number of polarized electrons in Earth due to the presence of Earth's geomagnetic field \cite{Hunter:2013hza}. Here, the Earth is treated as a polarized source and the Sun is treated as an unpolarized object. The ALP has a pseudoscalar coupling with the electrons in the Earth and scalar coupling with the nucleons in the Sun. This can give rise to an axion mediated monopole-dipole potential for an Earth-Sun system. We obtain constraints on monopole-dipole interaction strength in this pure astrophysical scenario from perihelion precession of Earth, gravitational light bending and Shapiro time delay. The bounds on the monopole-dipole coupling obtained from these gravity tests are strictly valid for the range of the force greater than the Earth-Sun distance which corresponds to the mass of the axion $\lesssim 10^{-18}~\rm{eV}$.

We also consider the axion mediated monopole-monopole potential between the unpolarized nucleons in the Earth and the Sun that can similarly affect the perihelion precession of planets, gravitational light bending and Shapiro time delay. We obtain constraints on monopole coupling from these tests of gravity. We also obtain constraints on dipole coupling from the excessive energy loss of the red giant branch. Multiplying these two couplings obtained from two different astrophysical observations, we obtain combined constraints on monopole-dipole coupling strength.

The paper is organized as follows. In Section \ref{sec2}, we discuss how the Earth can be treated as a polarized source. In Section \ref{sec3} we obtain the contribution of monopole dipole potential in perihelion precession of planets, gravitational light bending, and Shapiro time delay. We also obtain bounds on monopole-dipole strength from these tests of gravity with single astrophysical observation. In Section \ref{new}, we obtain constraints on monopole-dipole coupling strength from two different astrophysical observations. Finally, in Section \ref{sec4} we conclude and discuss our results. 

We use the natural system of units $(c=\hbar=1$, where $c$ is the speed of light and $\hbar$ is the reduced Planck constant) in our paper. We also choose Newton's gravitational constant $G=1$.   
\section{Earth as a polarized source}
\label{sec2}
\begin{figure}[h]
\includegraphics[height=4cm]{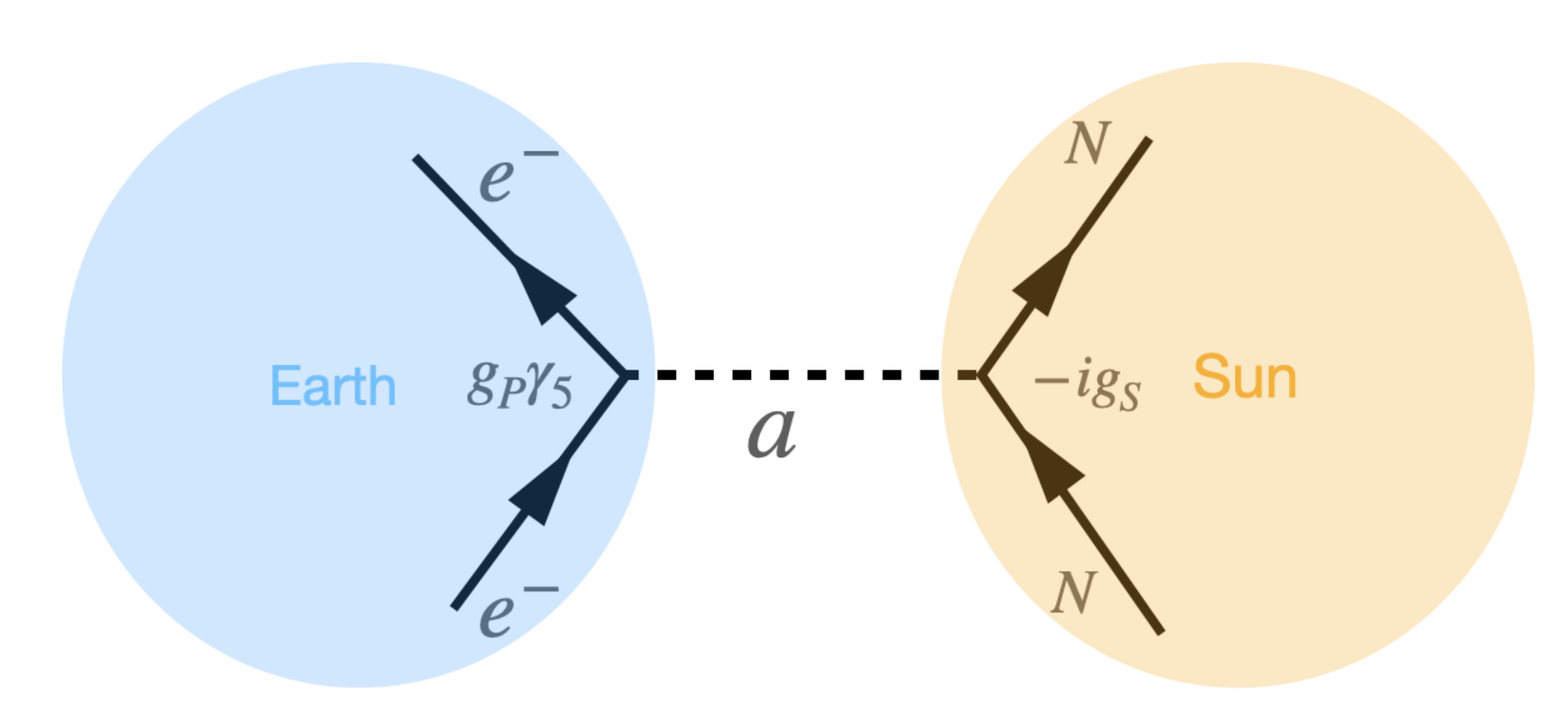}
\caption{Feynman diagram of $e^{-}N$ scattering mediated by ultralight ALP in an Earth-Sun system.}
\label{plot1}
\end{figure}
Recently, a long range dipole-dipole interaction arising between two spin polarized bodies is studied where the authors have considered the Earth as a source of spin polarized electrons \cite{Hunter:2013hza}. In the presence of the geomagnetic field, some of the electrons in paramagnetic minerals within the Earth acquire a small spin polarization. The magnitude and direction of the induced geoelectron spins depend on the Earth's material composition, geomagnetic field and temperature profile \cite{maus2010us}. The core of the Earth is mostly made of Fe-Ni alloy which does not contain any unpaired electron spins due to high pressure and temperature \cite{alfe1998first,alfe1998}. Hence, the Earth's core does not make any contribution to its polarization. The dominant contribution to the polarization comes from Fe, the most abundant transition metal in various oxides and silicates in the Earth's mantle and crust. Other major rock forming elements like Mg, Si, Al, and O have a negligible contribution to the Earth's polarization due to their closed electron shells. In \cite{Hunter:2013hza}, the electron spin density as a function of depth and all the mineral proportions in Earth's crust and mantle are mentioned very accurately. It is found that the unpaired electron density around $10^4~\rm{km}$ depth is about $10^{22}~/\rm{cm^3}$. Hence, the total unpaired electron spins inside the Earth will be $N_e\sim 10^{22}\times 10^{27}=10^{49}$. Most of the unpaired electrons exist in the $Fe^{2+}$ state with a total spin $s=2$, the so called $HS$ state. When the spin-$\frac{1}{2}$ electron in $HS$ $Fe^{2+}$ interacts with the external geomagnetic field, the spins become polarized and the polarization fraction becomes $\alpha=\frac{2\mu_B B}{kT}$, where the electron Bohr magneton is $\mu_B=\frac{e}{2m_e}=2.94\times 10^{-7}~\rm{eV^{-1}}$, $k$ is the Boltzmann constant, $B\sim 1~\rm{G}$ is the Earth's magnetic field in the mantle, and $T\sim 2000~\rm{K}$ is the temperature. Hence, we can obtain the polarization fraction as $\alpha\sim 10^{-7}$. Therefore, the total polarized electron spins in Earth is $N_e\times \alpha\sim 10^{49}\times 10^{-7}=10^{42}$. The value of $\alpha$ due to Earth's magnetic field is much larger than the accidental polarization, estimated as $\alpha_{\rm{accidental}}\sim\frac{1}{\sqrt{N_e}}\sim 10^{-25}$. These spin polarized geoelectrons can induce a net polarization due to Earth's magnetic field which can generate an axion mediated monopole-dipole potential for an Earth-Sun system. Such an interaction can affect the perihelion precession of Earth, gravitational light bending, and Shapiro time delay. However, the contribution of monopole-dipole potential for these observations is limited to be no larger than the measurement uncertainty. In FIG. \ref{plot1} we obtain the Feynman diagram for $e^{-}N$ scattering mediated by ultralight ALP for an Earth-Sun system. The ALP is coupled with the electrons in Earth by a pseudoscalar coupling. The ALP is also coupled with unpolarized nucleons in Sun by a scalar coupling. In the following, we obtain the contribution of monopole-dipole potential from the measurements of perihelion precession of Earth, gravitational light bending, and Shapiro time delay.     
\section{Perihelion precession, gravitational light bending and Shapiro time delay in presence of a monopole-dipole potential}
\label{sec3}
The success of Einstein's General Relativity (GR) theory has been consolidated by the observation of the perihelion precession of the Mercury planet. While orbiting around the Sun, the perihelion position of the Mercury planet shifts by a very small angle in each revolution. The dominant contribution to the perihelion shift comes from the gravitational effect of other solar bodies. There is also a subdominant contribution on perihelion shift due to the oblateness of Sun and Lense-Thirring precession. These non relativistic contributions are calculated based on Newtonian mechanics which follows $\frac{1}{r^2}$ force law. However, there is about $42.9799~\rm{arcsecond/century}$ \cite{Shapiro1990,Park_2017} mismatch from the observation after including all the non relativistic effects in the measurement of perihelion precession of Mercury. Einstein's general relativistic calculation of perihelion precession can completely resolve this anomaly. Besides Mercury, all the other planets also experience perihelion shifts. For example, the Earth has a perihelion shift of $3.84~\rm{arcsecond/century}$ due to GR correction. Since, Earth is taken as a polarized source, there can be an axion mediated monopole-dipole potential for an Earth-Sun system that can contribute to the perihelion precession measurement of Earth. However, the contribution of monopole-dipole interaction should be limited to be no larger than the measurement uncertainty which is $10^{-4}$ \cite{Pitjeva2005,Biswas:2008cw} for the Earth-Sun system. Using perturbative method, we analytically obtain the contribution of monopole-dipole potential in perihelion shift as (see Appendix~\ref{app2})  
\begin{equation}
\Delta\phi_{\rm{monopole-dipole}}\simeq\frac{g_S g_P N_1N_2}{2MD(1-\epsilon^2)M_P m_e}+\frac{g_Sg_P N_1N_2D^2m^3_a(1-\epsilon^2)}{6M_PM(1+\epsilon)m_e}+\mathcal{O}\Big((g_Sg_P)^2, m^4_a\Big).
\label{test1}
\end{equation}
Using the values of the solar mass $M=1.11\times 10^{57}~\rm{GeV}$, the Sun-Earth distance $D=0.98~\rm{AU}=7.37\times 10^{26}~\rm{GeV^{-1}}$, the eccentrcity of the Earth-Sun orbit $\epsilon=0.017$, the mass of the electron $m_e=5.1\times 10^{-4}~\rm{GeV}$, the Newton's gravitation constant $G=10^{-38}~\rm{GeV^{-2}}$, the mass of the planet Earth $M_P=3.35\times 10^{51}~\rm{GeV}$, the number of polarized electrons in Earth $N_1=10^{42}$, the number of unpolarized nucleons in the Sun $N_2=10^{57}$, we obtain the upper bound on monopole-dipole coupling as $g_Sg_P\lesssim 1.75\times 10^{-16}$ for mass of the axion $m_a\lesssim 1.35\times 10^{-18}~\rm{eV}$. We obtain this bound by considering that the contribution of monopole-dipole potential is limited to be no larger than the perihelion precession measurement uncertainty.

Besides, the perihelion precession of planets, gravitational light bending is another test of Einstein's GR theory \cite{Will:2014kxa,Will:2014zpa}. When a light ray from a distant pulsar comes to Earth, then the presence of a massive object like the Sun can distort the spacetime between the light source and the Earth. The increased gravitational potential due to the presence of the Sun decreases the speed of light and the light bends. The amount of bending depends on the mass of the gravitating object (Sun) and the impact parameter. In 1915, Einstein first calculated the amount of light bending due to the presence of the Sun based on the Equivalence principle. The calculated value of light bending is $1.75~\rm{arcsecond}$ which matches well with the experiment to an uncertainty of $10^{-4}$ \cite{Fomalont:2009zg}. The contribution of monopole-dipole potential should be limited to this uncertainty. We perturbatively calculate the contribution of monopole-dipole potential in gravitational light bending as (see Appendix~\ref{app3})
\begin{equation}
\begin{split}
\Delta\phi_{\rm{monopole-dipole}}\simeq -\frac{2m^3_ab^3g_Sg_PN_1N_2 \ln 2}{3M_PL^2\times 4\pi m_e}+\frac{g_Sg_PN_1N_2}{M_PL^2\times 4\pi m_e}\times \frac{4M}{b}-\frac{m^3_ab^3g_Sg_PN_1N_2}{3M_PL^2\times 4\pi m_e}\times \frac{4M}{b}\\
+\mathcal{O}\Big((g_Sg_P)^2, m^4_a\Big).
\end{split}
\label{test2}
\end{equation} 
We use $L^2=MD(1-\epsilon^2)$, and the value of impact parameter $b$ as the solar radius $b\sim R_\odot= 6.96\times 10^8~\rm{m}=3.51\times 10^{24}~\rm{GeV^{-1}}$. The contribution of monopole-dipole potential in the measurement of gravitational light bending should be within the measurement uncertainty and we obtain the bound on coupling as $g_Sg_P\lesssim 4.25\times 10^{-9}$ for $m_a\lesssim 1.35\times 10^{-18}~\rm{eV}$.

We also obtain constraints on monopole-dipole interaction strength from the Shapiro time delay. When a radar signal is sent from Earth to Venus and it reflects from Venus to Earth, then in this round trip, there is a time delay in getting the signal compared to the expectation. In 1964, Irwin Shapiro calculated the amount of time delay as $2\times 10^{-4}~\rm{s}$ \cite{Shapiro:1964uw,Shapiro:1968zza} which agrees well with the experiment to an uncertainty of $10^{-5}$ \cite{Bertotti:2003rm}. This time delay occurs due to the presence of strong gravitational potential near the Sun. The presence of long range monopole-dipole potential can contribute to the Shapiro time delay. However, its contribution should be within the measurement uncertainty. We analytically calculate the contribution of monopole-dipole potential in Shapiro time delay as (see Appendix \ref{app4})
\begin{equation}
\begin{split}
\Delta T_{\rm{monopole-dipole}}\simeq\frac{8M}{M_Pr_0E^2}\Big(m_a+\frac{1}{r_0}\Big)e^{-m_ar_0}\Big(\frac{g_Sg_PN_1N_2}{4\pi m_e}\Big)-\frac{4M}{M_PE^2r^2_0}\Big(\frac{g_Sg_PN_1N_2}{4\pi m_e}\Big)+\\
\mathcal{O}\Big((g_Sg_P)^2, m^2_a, M^2\Big).
\end{split}
\label{test3}
\end{equation}
Using the Earth-Sun distance $r_e=D=1.46\times 10^{11}~\rm{m}=7.37\times 10^{26}~\rm{GeV^{-1}}$, the Venus-Earth distance $r_v=1.08\times 10^{11}~\rm{m}=5.47\times 10^{26}~\rm{GeV^{-1}}$, the solar radius $r_0=R_\odot=6.96\times 10^8~\rm{m}=3.51\times 10^{24}~\rm{GeV^{-1}}$, and $E^2\simeq\frac{L^2}{r^2_0}\Big(1-\frac{2M}{r_0}\Big)$ we obtain the upper bound on coupling as $g_Sg_P\lesssim 1.08\times 10^{-4}$ for $m_a\lesssim 1.35\times 10^{-18}~\rm{eV}$. There is an extra multiplicative factor of $exp[-\frac{m^2_aL^2}{M}]$ in Eq. \ref{test1}, Eq. \ref{test2}, and Eq. \ref{test3} to incorporate the exponential suppression due to large value of axion mass.
\begin{figure}[h]
\includegraphics[height=6cm]{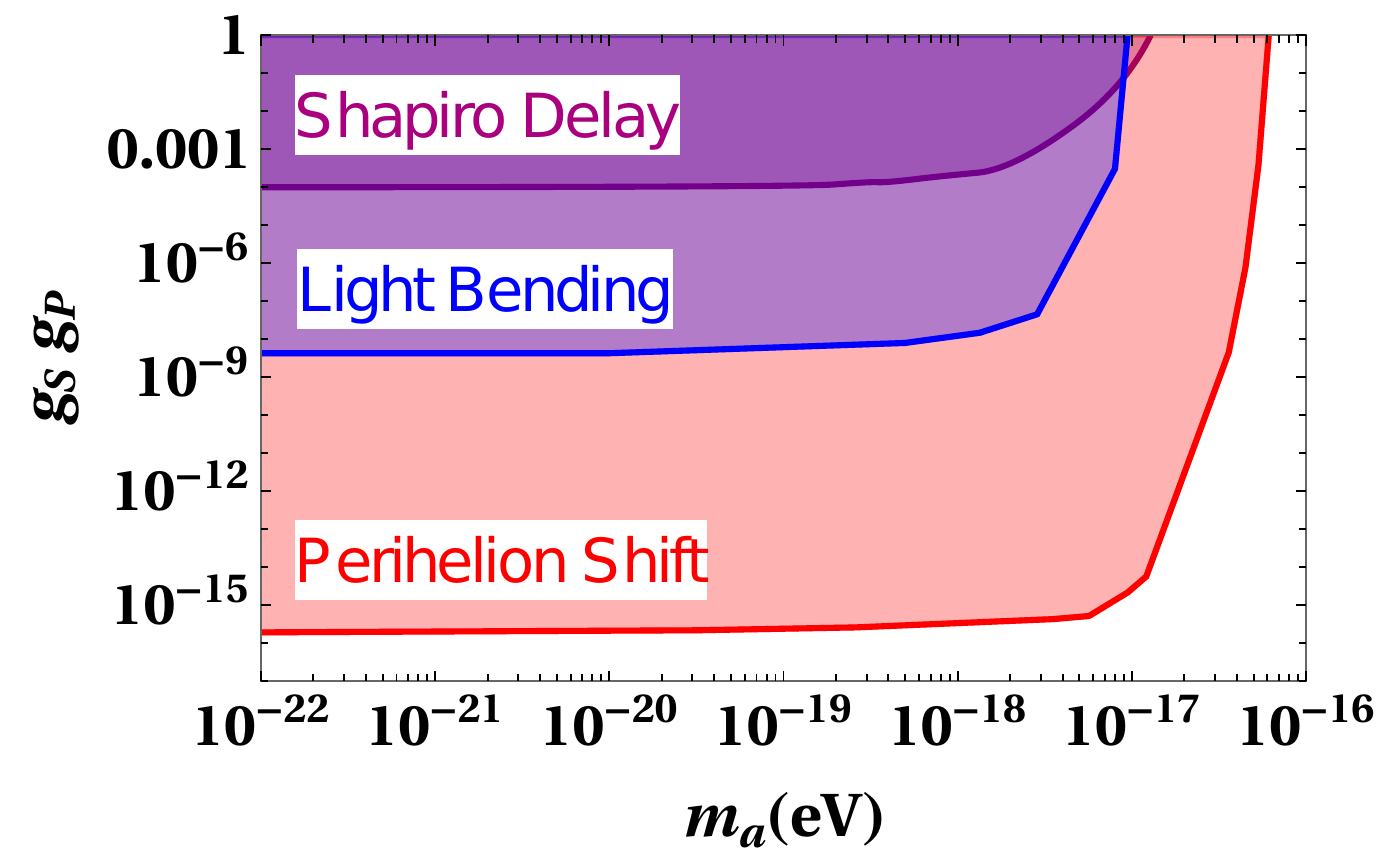}
\caption{Bounds on monopole-dipole interaction strength from single astrophysical observation}
\label{plot2}
\end{figure} 
In FIG. \ref{plot2} we obtain numerically the bounds on monopole-dipole coupling from perihelion precession of planets (red region), gravitational light bending (blue region), and Shapiro time delay (purple region). The shaded regions are excluded. We obtain stronger bound on $g_Sg_P$ from perihelion precession of planets as $g_Sg_P\lesssim 1.75\times 10^{-16}$ for the axions of mass $m_a\lesssim 1.35\times 10^{-18}~\rm{eV}$. This is the first bound on $g_Sg_P$ that we obtain from a single astrophysical observation and for ALPs of mass $m_a\lesssim \mathcal{O}(10^{-18})~\rm{eV}$.
\section{Constraints on monopole-dipole coupling from two different astrophysical observations}\label{new}
\begin{figure}[h]
\includegraphics[height=4cm]{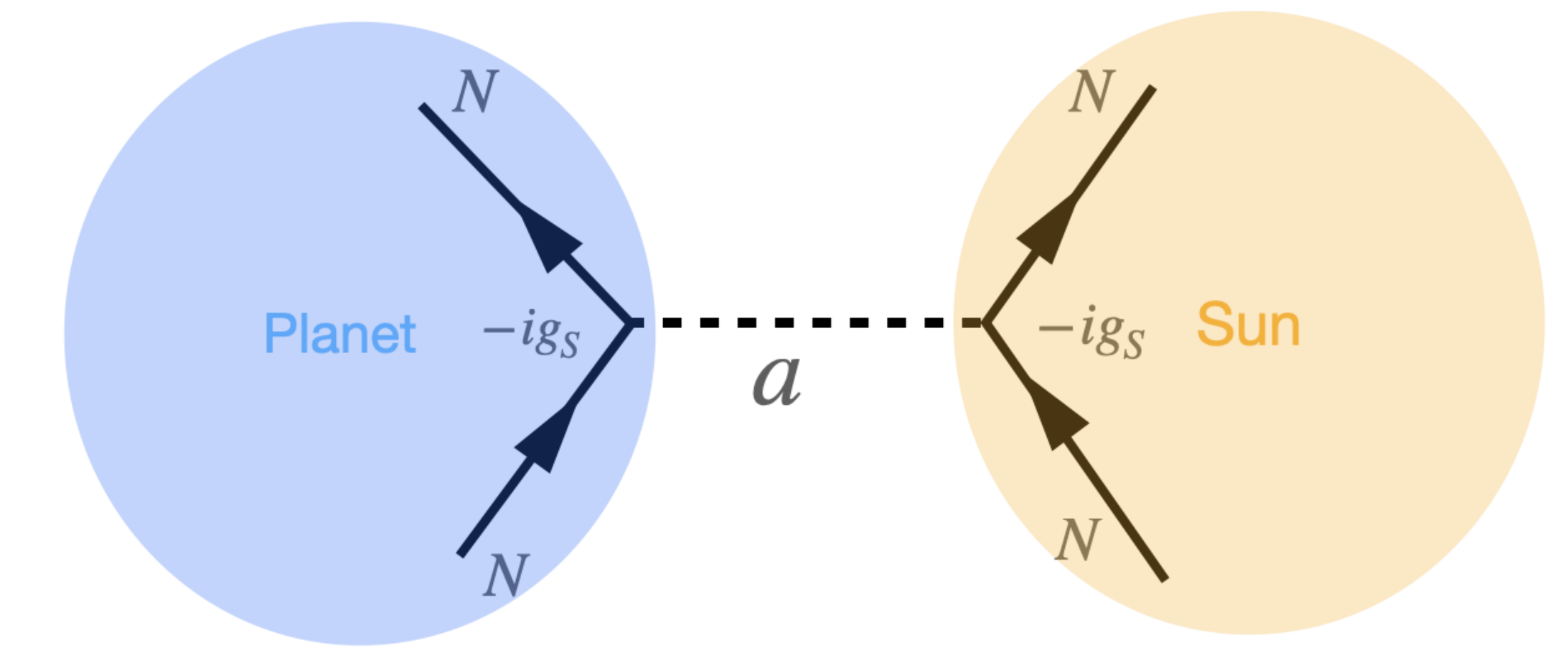}
\caption{Monopole coupling of axions with nucleons in the Earth and the Sun}
\label{plotp}
\end{figure} 
In this section, we obtain constraints on monopole-dipole coupling from two different astrophysical observations. In FIG. \ref{plotp}, we consider monopole-monopole coupling of axions with unpolarized nucleons in the planet and the Sun that can change the perihelion precession of planets, gravitational light bending and Shapiro time delay within the measurement uncertainty. The potential due to axion mediated nucleon-nucleon scattering in the Earth-Planet system is $\frac{g^2_SN_1N_2}{4\pi r}e^{-m_ar}$, where $N_1$ and $N_2$ are the numbers of nucleons in the Sun and the planet respectively. Hence, the perihelion shift due to the axion mediated monopole-monopole potential between the Sun and the planet is \cite{KumarPoddar:2020kdz}
\begin{equation}
\Delta\phi_{\rm{monopole-monopole}}\simeq \frac{g^2_S N_1N_2m^2_a D^2(1-\epsilon^2)}{4M_P (M+\frac{g^2_SN_1N_2}{4\pi M_P})(1+\epsilon)}+\mathcal{O}(g^3_S, m^3_a),
\label{re1}
\end{equation}
where $M_P$ is the mass of the planet, $M$ is the mass of the Sun, and $D$ is the semi-major axis of the planetary orbit with eccentricity $\epsilon$. The contribution of axion mediated monopole-monopole potential should be limited to be no larger than the perihelion precession measurement uncertainty. We obtain the stronger bound on $g_S$ for the planet Mars \cite{KumarPoddar:2020kdz} and its value is $g_S\lesssim 3.51\times 10^{-25}$ for the mass of the axion $m_a\lesssim 1.35\times 10^{-18}~\rm{eV}$.

The bending of light due to the axion mediated monopole potential is \cite{Poddar:2021sbc}
\begin{equation}
\Delta \phi _{\rm{monopole-monopole}}\simeq \frac{g^2_SN_1N_2b}{2\pi M_PL^2}(1-0.347m^2_ab^2)-\frac{g^2_SN_1N_2Mm^2_ab^2}{2\pi M_PL^2}+\mathcal{O}(g^3_S, m^3_a).
\label{re2}
\end{equation}
We obtain the constraint on axion monopole coupling from the gravitational light bending as $g_S\lesssim 5.82\times 10^{-23}$ for the axions of mass $m_a\lesssim 1.35\times 10^{-18}~\rm{eV}$.

Similarly, the contribution of axion mediated monopole potential in Shapiro time delay is \cite{Poddar:2021sbc}
\begin{equation}
\begin{split}
\Delta T_{\rm{monopole-monopole}}\simeq 2b_0c_0(-1+c_0M)(r_e+r_v)+\frac{b_0c^2_0}{2}(r^2_e+r^2_v)+2b_0-4c_0Mb_0+\\
2a_0(r_e+r_v)+\frac{b_0}{24}(48+36c^2_0r^2_0[E_i(-c_0r_e)+E_i(-c_0r_v)])+\mathcal{O}(g^3_S, m_a^3),
\end{split}
\label{re3}
\end{equation}
where $a_0=\frac{g^2_SN_1N_2e^{-m_ar_0}}{4\pi M_PE^2r_0}$, $b_0=\frac{g^2_SN_1N_2}{4\pi M_P E^2}$, and $c_0=m_a$.

We obtain the constraint on axion monopole coupling from the Shapiro time delay as $g_S\lesssim 3.59\times 10^{-22}$ for the axion mass $m_a\lesssim 1.35\times 10^{-18}~\rm{eV}$. There is an extra multiplicative factor of $exp[-\frac{m^2_aL^2}{M}]$ in Eq. \ref{re1}, Eq. \ref{re2} and Eq. \ref{re3} if we solve the perihelion shift, light bending and Shapiro time delay numerically for the axion mediated monopole-monopole potential to incorporate the exponential suppression due to large values of axion mass.

The bound on the axion electron pseudoscalar coupling can be obtained from the cooling of red giant stars and white dwarfs. The axion electron coupling allows the stellar energy loss by the bremsstrahlung $(e+Ze\rightarrow e+Ze+a)$ and Compton process $(\gamma+e\rightarrow e+a)$ \cite{Raffelt:1999tx,Raffelt:2006cw}. The excessive energy loss due to these processes will delay the Helium ignition in the red giant stars. Therefore the tip of the red giant branch becomes brighter. The measurement of the tip of the red giant branch in the $\omega$ Centaury from Gaia DR2 data put bound on the axion -electron coupling as $g_P\lesssim 1.6\times 10^{-13}$ for the mass of the axions $m_a\lesssim 10~\rm{keV}$ \cite{Capozzi:2020cbu}.
\begin{figure}[h]
\includegraphics[height=6cm]{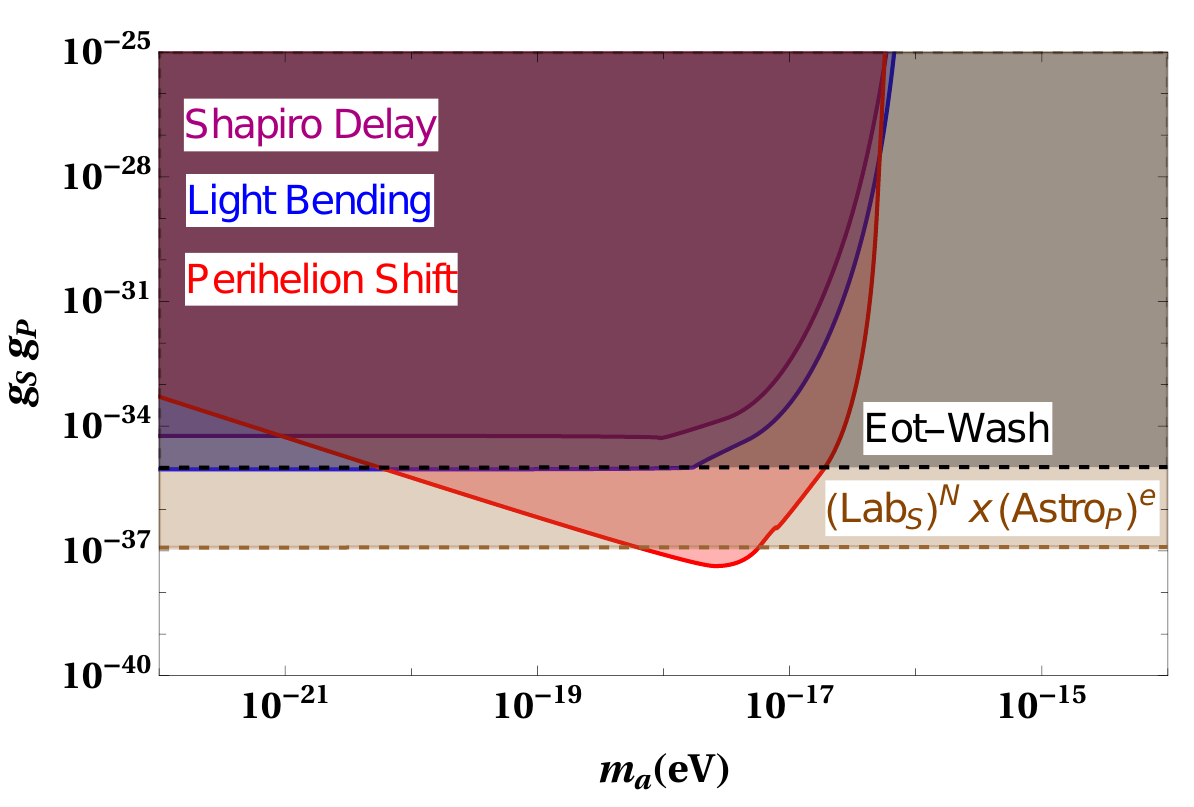}
\caption{Bounds on monopole-dipole interaction strength from two different astrophysical observations}
\label{nt}
\end{figure} 

To obtain the bound on monopole-dipole coupling $(g_Sg_P)$, we take the product of the bounds on $g_S$ obtained from the tests of gravity (perihelion precession of planet, gravitational light bending and Shapiro time delay) and $g_P$ obtained from the energy loss from the red giant branch. In FIG.\ref{nt} we obtain the bounds on $g_Sg_P$ from two different astrophysical observations. The perihelion precession of planets and red giant branch give the bound on monopole-dipole coupling as $g_Sg_P\lesssim 5.61\times 10^{-38}$. We also obtain the bound on $g_Sg_P$ from gravitational light bending and red giant branch as $g_Sg_P\lesssim 9.31\times 10^{-36}$. Lastly, the bound on $g_Sg_P$ obtained from Shapiro time delay and red giant branch is $g_Sg_P\lesssim 5.74\times 10^{-35}$. These bounds are only valid for the mass of the axion $m_a\lesssim 1.35\times 10^{-18}~\rm{eV}$. We obtain the stronger bound on $g_Sg_P$ from the perihelion precession of planets and energy loss of the red giant branch. The shaded regions in FIG. \ref{nt} are excluded. The bound $g_Sg_P\lesssim 5.61\times 10^{-38}$ is three orders of magnitude stronger than the Eot-Wash experiment \cite{Berge:2017ovy} and one order of magnitude stronger than the $(\rm{Lab})^N_S\times (\rm{Astro})^e_P$ limit \cite{OHare:2020wah}.
\section{Conclusions and Discussions}
\label{sec4}
In this paper, we obtain constraints on monopole-dipole coupling strength from single astrophysical observations such as the perihelion precession of Earth, gravitational light bending, and Shapiro time delay. These bounds are strictly valid for the ALP mass $m_a\lesssim 1.35\times 10^{-18}~\rm{eV}$. Due to the presence of a geomagnetic field, $10^{42}$ number of electrons can be polarized in Earth and ALP mediated monopole-dipole force can act between the Earth and the Sun. We obtain a stronger bound on monopole-dipole coupling strength from perihelion precession of the planet as $g_Sg_P\lesssim 1.75\times 10^{-16}$ from single astrophysical observation.

The previous lab-astro bounds on $g_Sg_P$ obtained in the literature are derived from two different observations. In these studies, the monopole coupling $g_S$ and the dipole coupling $g_P$ are measured independently from two different observations and they are simply multiplied to get bound on $g_Sg_P$. To get the bound in this way is overly stringent and may not be completely reliable if the axion changes its behaviour in different environments. The bounds on $g_Sg_P$ obtained from lab-lab experiments are only valid for the axions of mass $\rm{\mu eV}\lesssim m_a\lesssim \rm{meV}$.

In this work, we have obtained the first bounds on $g_Sg_P$ from single astrophysical observations. In all of these observations, the massless limit gives the stronger bound on monopole-dipole coupling strength. In the massless limit, the perihelion shift is inversely proportional to the Sun-planet distance. This means planets which are closer to Sun will put the best bound on $g_Sg_P$. However, to achieve an improved bound on $g_Sg_P$ from perihelion precession, one needs to calculate accurately the number of polarized spins in those planets.

The bounds on the monopole-dipole couplings that we obtain from perihelion precession, gravitational light bending, and Shapiro time delay are the order of magnitude calculations. These bounds strongly depend on the number of polarized electrons in the Earth which is not a fixed quantity at all its layers. In fact, this number depends on the magnetic field and temperature at each layer of the Earth which varies with its depth. Hence, at the massless limit, the monopole-dipole coupling strength will not be a fixed quantity and it should have different values at different depths. We obtain the number of polarized electrons in Earth as $10^{42}$ by taking the average values of Earth's magnetic field, temperature and the number density of unpaired electrons which we have taken as fixed quantities. Therefore, our bounds on monopole-dipole couplings are constant at the massless limits. This is the first study to probe monopole-dipole coupling from single astrophysical observations for $m_a\lesssim 1.35\times 10^{-18}~\rm{eV}$. Our bounds on the monopole-dipole coupling are the order of magnitude calculation and can be significantly improved by accurate incorporation of the number of polarized spins at each layer of Earth from geochemical and geological surveys. Such analyses are important to probe these long range spin dependent interactions.    

We also obtain constraints on monopole-dipole coupling strength from two different astrophysical observations. We consider monopole coupling of axions with unpolarized nucleons in the Earth and the Sun to obtain bounds on monopole coupling from perihelion precession of planets, gravitational light bending and Shapiro time delay. Multiplying these monopole couplings with the dipole coupling obtained from excessive energy loss of the red giant branch, we derive the monopole-dipole coupling strength. For $m_a\lesssim 1.35\times 10^{-18}~\rm{eV}$, we obtain $g_Sg_P\lesssim 5.61\times 10^{-38}$ from perihelion precession and red giant branch which is three orders of magnitude stronger than the Eot-Wash experiment and one order of magnitude stronger than the current $(\rm{Lab})^N_S\times(\rm{Astro})^e_P$ limit.

We can also constrain the axion mediated monopole-dipole coupling between nucleonic currents. The cooling of hot neutron star HESS J1731-347 puts bound on axion nucleon pseudoscalar coupling as $g^N_P\lesssim 2.8\times 10^{-10}$. We also obtain axion nucleon scalar coupling as $g^N_S\lesssim 3.51\times 10^{-25}$. Combining these two couplings, we obtain the bound on the monopole-dipole coupling strength for only nucleonic currents as $g^N_Sg^N_P\lesssim 9.83\times 10^{-35}$ for the mass of the axions $m_a\lesssim 1.35\times 10^{-18}~\rm{eV}$. This bound is better than the projected ARIADNE experiment \cite{Arvanitaki:2014dfa} and $(\rm{Lab})^N_S\times (\rm{Astro})^N_P$ by a factor of $2$ \cite{OHare:2020wah}. Future space missions with better precision can significantly improve the bounds of monopole-dipole couplings. These ultralight axions can be promising candidates for fuzzy dark matter.
\section*{Acknowledgements}
The authors would like to thank Srubabati Goswami for useful discussions in the initial part of this work. The authors would also like to thank Subhendra Mohanty for reading the manuscript. T.K.P is indebted to Basudeb Dasgupta and Ranjan Laha for fruitful discussions and suggestions.   
\appendix
\section{Monopole-dipole potential due to polarized electron and unpolarized nucleon scattering}\label{app1}
In FIG. \ref{feynman} we show the Feynman diagram of $e^-N$ scattering mediated by pseudoscalar ALP $(a)$. The axion is coupled to the polarized electron with coupling constant $g_P$ and to the unpolarized nucleon with coupling constant $g_S$. Hence, the amplitude of the above process becomes
\begin{eqnarray}
i\mathcal{M}&=&\bar{u}_{s^\prime_1}(p^\prime_1)g_P \gamma_5 u_{s_1}(p_1)\frac{i}{q^2-m^2_a}\bar{u}_{s^\prime_2}(p^\prime_2)(-ig_S)u_{s_2}(p_2)\nonumber\\
&=& \frac{g_Pg_S}{q^2-m^2_a}\bar{u}_{s^\prime_1}(p^\prime_1)\gamma_5 u_{s_1}(p_1)\bar{u}_{s^\prime_2}(p^\prime_2)u_{s_2}(p_2),
\label{eq:1}
\end{eqnarray}
where $q=p_1-p^\prime_1=p^\prime_2-p_2$. In the Non Relativistic (NR) limit, all three momentum components are much smaller than the mass of the particle $(m)$ and hence, the energy of the particle is $E\approx m$. We also choose the normalization condition $u^\dagger_{s^\prime}(p)u_s(p)=\delta_{ss^\prime}$. We can also write the positive energy spinor in the NR limit as
\begin{equation}
u_s(p)=\Big(1-\frac{\gamma_ip_i}{2m}\Big)\chi_s+\mathcal{O}(p^2),
\label{eq:2}
\end{equation} 
where $\chi_s$ is a normalized eigenvector satisfying $\chi^\dagger_s\gamma^0=\chi^\dagger_s$ and $\gamma_0\chi_s=\chi_s$. Here, $\gamma_i$ denotes the Dirac gamma matrices and $i$ runs from $1$ to $3$. Hence, in the NR limit, we can calculate the following bilinear terms using Eq. \ref{eq:2} as
\begin{equation}
\bar{u}_{s^\prime_2}(p^\prime_2)u_{s_2}(p_2)=1, \hspace{0.6cm} \bar{u}_{s^\prime_1}(p^\prime_1)\gamma_5 u_{s_1}(p_1)=\frac{1}{2m_e}\chi^\dagger_{s^\prime_1} \mathbf{\sigma}.\mathbf{q} \chi_{s_1},
\label{eq:3}
\end{equation}
where $\mathbf{\sigma}$ denotes the Pauli spin vector and $m_e$ denotes the mass of the polarized electron. We can write the amplitude (Eq. \ref{eq:1}) of $e^-N$ scattering process as 
\begin{equation}
\mathcal{M}=\frac{ig_Pg_S}{|\mathbf{q}|^2+m^2_a}\bar{u}_{s^\prime_1}(p^\prime_1)\gamma_5u_{s_1}(p_1)\bar{u}_{s^\prime_2}(p^\prime_2)u_{s_2}(p_2),
\label{eq:4}
\end{equation}
where we can write $q^2={q^0}^2-|\mathbf{q}|^2$, and $|q^0|\ll |\mathbf{q}|$ in the NR limit. Using, Eq. \ref{eq:3} we can write the potential for $e^-N$ scattering as
\begin{equation}
V(r)=-\int \frac{d^3q}{(2\pi)^3}e^{i\mathbf{q}.\mathbf{r}}\Big(\frac{ig_Pg_S}{|\mathbf{q}|^2+m^2_a}\Big)\frac{\mathbf{s_1}.\mathbf{q}}{m_e},
\label{eq:5}
\end{equation}
where the spin vector is $\mathbf{s_1}=\frac{\mathbf{\sigma}}{2}$. Therefore, the potential becomes 
\begin{eqnarray}
V(r)&=&-\frac{g_Pg_S}{m_e}(\mathbf{s_1}.\mathbf{\nabla})\int \frac{d^3q}{(2\pi)^3}\frac{1}{|\mathbf{q}|^2+m^2_a}e^{i\mathbf{q}.\mathbf{r}}\nonumber\\
&=& -\frac{g_Pg_S}{m_e}(\mathbf{s_1}.\mathbf{\nabla})\frac{1}{4\pi r}e^{-m_ar}\nonumber\\
&=& \frac{g_Pg_S}{4\pi m_e}(\mathbf{s_1}.\hat{r})\Big(\frac{m_a}{r}+\frac{1}{r^2}\Big)e^{-m_a r}.
\label{eq:5}
\end{eqnarray}
\bibliographystyle{utphys}
This is the expression for monopole-dipole potential which can act between a polarized and an unpolarized objects.
\section{Perihelion precession of Earth in presence of a long range monopole-dipole potential}\label{app2}
If Earth contains polarized electrons then long range monopole-dipole potential can act between the Earth and the Sun. This new long range force mediated by ultralight ALP can contribute to the perihelion precession of Earth. However, its contribution is limited to be no larger than the uncertainty in the measurement of perihelion precession. For a timelike particle, we can write $g_{\mu\nu}\dot{x}^\mu\dot{x}^\nu=-1$, where $g_{\mu\nu}$ is the metric tensor for the Schwarzschild background spacetime. In presence of a long range monopole-dipole potential, we can write
\begin{equation}
\frac{E^2-1}{2}=\frac{\dot{r}^2}{2}+\frac{L^2}{2r^2}-\frac{ML^2}{r^3}-\frac{M}{r}-\frac{\beta Em_a}{M_Pr}e^{-m_ar}-\frac{\beta E}{M_P r^2}e^{-m_ar},
\label{eq:6a}
\end{equation}   
where $\dot{r}=\frac{L}{r^2}\frac{dr}{d\phi}$, $M$ and $M_P$ are the masses of the Sun and Earth respectively, and $\beta=\frac{g_Sg_PN_1N_2}{4\pi m_e}$. $N_1$ and $N_2$ are the numbers of polarized electrons in the Earth and unpolarized nucleons in the Sun respectively. We have also neglected the $\mathcal{O}(\beta^2)$ term because the coupling for the monopole-dipole potential is small. $E$ is a constant of motion which is termed as the total energy per unit rest mass for a timelike geodesic relative to an observer in rest frame at infinity. The total energy of the system per unit mass for a very small eccentric orbit in presence of a monopole-dipole potential is 
\begin{equation}
E\approx 1-\frac{M}{2D}-\frac{g_Sg_PN_1N_2}{4\pi m_e}e^{-m_a D}\Big(\frac{m_a}{M_PD}+\frac{m^2_a}{2M_P}+\frac{1}{M_P D^2}\Big),
\label{energy}
\end{equation}
and $L$ is another constant of motion which is the angular momentum per unit mass of the system. In Eq. \ref{eq:6a}, the first term on the right hand side denotes the kinetic energy part, the second term denotes the centrifugal potential part, the third term arises due to the contribution of GR, the fourth term denotes the Newtonian potential, and the last two terms appear due to the contribution of monopole-dipole potential. We can write Eq. \ref{eq:6a} in terms of reciprocal coordinate $u=\frac{1}{r}$ as
\begin{equation}
\Big(\frac{du}{d\phi}\Big)^2+u^2=\frac{E^2-1}{L^2}+2Mu^3+\frac{2Mu}{L^2}+\frac{2\beta Em_au}{L^2M_P}e^{-\frac{m_a}{u}}+\frac{2\beta Eu^2}{L^2M_P}e^{-\frac{m_a}{u}},
\label{eq:6}
\end{equation} 
where $\phi$ denotes the azimuthal coordinate. Expanding the exponential term in Eq. \ref{eq:6} and take derivative with respect to $\phi$, we obtain
\begin{equation}
\frac{d^2u}{d\phi^2}+u=\frac{M}{L^2}+3Mu^2+\frac{2\beta E u}{L^2 M_P}-\frac{\beta Em^3_a}{3L^2M_P u^2}.
\label{eq:7}
\end{equation} 
To solve this second order differential equation we consider $u=u_0(\phi)+\Delta u(\phi)$, where $u_0(\phi)$ is the solution for Newton's theory and $\Delta u(\phi)$ is the solution due to the contribution of GR and monopole-dipole potential. Hence, we can write
\begin{equation}
\frac{d^2u_0}{d\phi^2}+u_0=\frac{M}{L^2}.
\label{eq:8}
\end{equation}
The solution of Eq. \ref{eq:8} becomes
\begin{equation}
u_0(\phi)=\frac{M}{L^2}(1+\epsilon\cos\phi),
\label{eq:9}
\end{equation}
where $\epsilon$ is the eccentricity of the Earth-Sun elliptic orbit. The differential equation for $\Delta u$ is 
\begin{equation}
\frac{d^2\Delta u}{d\phi^2}+\Delta u=\frac{3M^3}{L^4}(1+\epsilon^2\cos^2\phi+2\epsilon\cos\phi)+\frac{2\beta M E}{L^4M_P}(1+\epsilon\cos\phi)-\frac{\beta E m^3_a L^2}{3M_P M^2(1+\epsilon^2\cos\phi+2\epsilon\cos\phi)}.
\label{eq:10}
\end{equation}
The solution of Eq. \ref{eq:10} becomes 
\begin{equation}
\Delta u=\frac{3M^3}{L^4}\epsilon\phi\sin\phi+\frac{\beta ME}{L^4M_P}\epsilon\phi\sin\phi+\frac{\beta E m^3_aL^2}{3M_P M^2}\frac{\epsilon\sin\phi}{(1-\epsilon^2)^\frac{3}{2}}\times \frac{\sqrt{1-\epsilon^2}}{(1+\epsilon)}\phi,
\label{eq:11}
\end{equation}
where we keep terms which are linear in $\phi$ and hence contribute to the perihelion precession of Earth. Hence, the total solution of Eq. \ref{eq:7} becomes
\begin{equation}
u=u_0(\phi)+\Delta u(\phi)=\frac{M}{L^2}(1+\epsilon\cos\phi)+\frac{3M^3}{L^4}\epsilon\phi\sin\phi+\frac{\beta ME}{L^4M_P}\epsilon\phi\sin\phi+\frac{\beta E m^3_a L^2}{3M_P M^2}\frac{\epsilon\sin\phi}{(1+\epsilon)(1-\epsilon^2)}.
\label{eq:12}
\end{equation} 
We can also write Eq. \ref{eq:12} as
\begin{equation}
u=\frac{M}{L^2}[1+\epsilon\cos\phi(1-\gamma)],
\label{eq:13}
\end{equation}
where 
\begin{equation}
\gamma=\frac{3M^2}{L^2}+\frac{\beta}{L^2M_P}+\frac{\beta L^4m^3_a}{3M_PM^3}\frac{1}{(1+\epsilon)(1-\epsilon^2)},
\label{eq:14}
\end{equation}
Here, we take $E\approx 1$ as other terms in Eq. \ref{energy} compared to $1$ are very small. Here, $D$ denotes the semi major axis of the orbit. As $\phi\rightarrow \phi+2\pi$, $u$ is not the same. Therefore, Earth does not follow its previous orbit. Hence, the change in the azimuthal angle or the perihelion shift becomes
\begin{equation}
\Delta\phi=\frac{2\pi}{1-\gamma}-2\pi=2\pi \gamma=\frac{6\pi M^2}{L^2}+\frac{2\pi\beta}{L^2M_P}+\frac{2\pi\beta L^4m^3_a}{3M_P M^3}\frac{1}{(1+\epsilon)(1-\epsilon^2)}.
\label{eq:15}
\end{equation}
Substituting $L^2=MD(1-\epsilon^2)$, and $\beta=\frac{g_Sg_PN_1N_2}{4\pi m_e}$, we obtain
\begin{equation}
\Delta\phi=\frac{6\pi  M}{D(1-\epsilon^2)}+\frac{g_S g_P N_1N_2}{2MD(1-\epsilon^2)M_P m_e}+\frac{g_Sg_P N_1N_2D^2m^3_a(1-\epsilon^2)}{6M_PM(1+\epsilon)m_e},
\label{eq:16}
\end{equation}
Eq. \ref{eq:16} is the general expression of perihelion shift due to monopole-dipole potential between a polarized object and an unpolarized object. The first term on the right hand side arises due to the GR contribution in perihelion shift and its value for Earth is $3.84~\rm{arcsecond/century}$. The last two terms arise due to the contribution of monopole-dipole potential. In $g_Sg_P\rightarrow 0$ limit, we get back the standard GR term. Hence, the contribution of monopole-dipole potential in perihelion shift is
\begin{equation}
\Delta\phi_{\rm{monopole-dipole}}\simeq\frac{g_S g_P N_1N_2}{2MD(1-\epsilon^2)M_P m_e}+\frac{g_Sg_P N_1N_2D^2m^3_a(1-\epsilon^2)}{6M_PM(1+\epsilon)m_e}+\mathcal{O}\Big((g_Sg_P)^2, m^4_a\Big).
\end{equation}
\section{Gravitational light bending in presence of a long range monopole-dipole potential}\label{app3}
Light follows the null geodesic which is given by
\begin{equation}
g_{\mu\nu}V^\mu V^\nu=0,
\label{eq:17}
\end{equation}
where $V^\mu=\frac{dx^\mu}{d\lambda}$ is the tangent vector along the path parametrized by $x^\mu(\lambda)$, where $\lambda$ is the affine parameter. For a Schwarzschild background and planar motion, the conserved quantities are $E=(1-\frac{2M}{r})\dot{t}$ and $L=r^2\dot{\phi}$. Here, $E$ and $L$ denote the total energy and angular momentum per unit mass of the system respectively. We can write the null geodesic in terms of these conserved quantities as
\begin{equation}
\frac{E^2}{2}=\frac{L^2}{2}\Big(\frac{du}{d\phi}\Big)^2+\frac{L^2u^2}{2}(1-2Mu),
\label{eq:18}
\end{equation}
where we use $\dot{r}=\frac{dr}{d\lambda}=\frac{L}{r^2}\frac{dr}{d\phi}$ and the reciprocal coordinate $u=\frac{1}{r}$. The presence of long range monopole-dipole potential changes the effective potential of the Sun-Earth system as
\begin{equation}
V_{\rm{eff}}=\frac{L^2}{2}\Big(\frac{du}{d\phi}\Big)^2+\frac{L^2u^2}{2}(1-2Mu)-\frac{\beta m_a u}{M_P}e^{-\frac{m_a}{u}}-\frac{\beta u^2}{M_P}e^{-\frac{m_a}{u}},
\label{eq:19}
\end{equation}
where the last two terms arise due to the presence of long range monopole-dipole potential. Hence, Eq. \ref{eq:18} becomes 
\begin{equation}
\frac{E^2}{2}=\frac{L^2}{2}\Big(\frac{du}{d\phi}\Big)^2+\frac{L^2u^2}{2}(1-2Mu)-\frac{\beta m_a u}{M_P}e^{-\frac{m_a}{u}}-\frac{\beta u^2}{M_P}e^{-\frac{m_a}{u}}.
\label{eq:20}
\end{equation} 
Differentiating Eq. \ref{eq:20} and expanding the exponential term we obtain
\begin{equation}
\frac{d^2u}{d\phi^2}+u=3Mu^2+\frac{2\beta u}{M_PL^2}-\frac{\beta m^3_a}{3u^2M_PL^2}
\label{eq:21}
\end{equation}
To solve this second order differential equation, we consider $u(\phi)=u_0(\phi)+\Delta u(\phi)$, where $u_0(\phi)$ is the solution for the complementary function and $\Delta u(\phi)$ is the solution for particular integral. Hence, we can write
\begin{equation}
\frac{d^2u_0}{d\phi^2}+u_0=0,
\label{eq:22}
\end{equation}
and the solution of Eq. \ref{eq:22} is $u_0=\frac{\sin\phi}{b}$, where $b$ is the impact parameter. To find the solution to a particular integral, we can write
\begin{equation}
\frac{d^2\Delta u}{d\phi^2}+\Delta u=\frac{3M\sin^2\phi}{b^2}+\frac{2\beta\sin\phi}{M_PL^2b}-\frac{\beta m^3_ab^2}{3M_PL^2\sin\phi}.
\label{eq:23}
\end{equation} 
The solution of Eq. \ref{eq:23} is
\begin{equation}
\Delta u(\phi)=\frac{3M}{2b^2}\Big(1+\frac{1}{3}\cos2\phi\Big)+\frac{2\beta}{M_PL^2b}\Big(-\frac{\phi\cos\phi}{2}\Big)-\frac{\beta m^3_a b^2}{3M_PL^2}[\cos\phi \ln|\csc\phi+\cot\phi|-1].
\label{eq:24}
\end{equation}
Hence, the total solution of Eq. \ref{eq:21} becomes
\begin{equation}
u(\phi)=\frac{\sin\phi}{b}+\frac{3M}{2b^2}\Big(1+\frac{1}{3}\cos2\phi\Big)-\frac{\beta\phi\cos\phi}{M_PL^2b}-\frac{\beta m^3_ab^2}{3M_PL^2}[\cos\phi\ln|\csc\phi+\cot\phi|-1].
\label{eq:25}
\end{equation}
At a far distance from the Sun, $u\rightarrow 0$ as $\phi\rightarrow 0$. Hence, we can write the change in the angular coordinate as
\begin{equation}
\delta \phi=\frac{-\frac{2M}{b^2}+\frac{\beta m^3_ab^2}{3M_PL^2}\ln 2}{\frac{1}{b}-\frac{\beta}{M_PL^2b}+\frac{\beta m^3_ab^2}{3M_PL^2}}.
\label{eq:26}
\end{equation}
From the symmetry argument, we can claim that the contribution to $\delta\phi$ before and after the turning points are the same. Therefore, the total light bending is
\begin{equation}
\Delta\phi=-2\delta\phi=\frac{\frac{4M}{b^2}-\frac{2\beta m^3_ab^2}{3M_PL^2}\ln 2}{\frac{1}{b}-\frac{\beta}{M_PL^2b}+\frac{\beta m^3_ab^2}{3M_PL^2}}.
\label{eq:27}
\end{equation}
In the $\beta\rightarrow 0$ limit, we obtain $\Delta\phi=\frac{4M}{b}$, which is the standard GR result for gravitational light bending. Hence, we can write the contribution of monopole-dipole potential in gravitational light bending is
\begin{equation}
\Delta\phi_{\rm{monopole-dipole}}=\frac{\frac{4M}{b^2}-\frac{2 m^3_ab^2\ln 2}{3M_PL^2}\frac{g_Sg_PN_1N_2}{4\pi m_e}}{\frac{1}{b}-\frac{1}{M_PL^2b}\frac{g_Sg_PN_1N_2}{4\pi m_e}+\frac{ m^3_ab^2}{3M_PL^2}\frac{g_Sg_PN_1N_2}{4\pi m_e}}-\frac{4M}{b}+\mathcal{O}\Big((g_Sg_P)^2, m^4_a\Big).
\label{eq:28}
\end{equation} 
We can also write Eq. \ref{eq:28} as 
\begin{equation}
\begin{split}
\Delta\phi_{\rm{monopole-dipole}}\simeq -\frac{2m^3_ab^3g_Sg_PN_1N_2 \ln 2}{3M_PL^2\times 4\pi m_e}+\frac{g_Sg_PN_1N_2}{M_PL^2\times 4\pi m_e}\times \frac{4M}{b}-\frac{m^3_ab^3g_Sg_PN_1N_2}{3M_PL^2\times 4\pi m_e}\times \frac{4M}{b}\\
+\mathcal{O}\Big((g_Sg_P)^2, m^4_a\Big).
\end{split}
\end{equation} 
\section{Shapiro time delay in presence of a long range monopole-dipole potential}\label{app4}
When a radar signal is sent from Earth to Venus and the signal reflects from Venus to Earth then due to the presence of the Sun between Earth and Venus there is a time delay in the round trip compared to the case if there is no Sun. We can write Eq. \ref{eq:20} as 
\begin{equation}
\frac{E^2}{2}=\frac{\dot{r}^2}{2}+\frac{L^2}{2r^2}\Big(1-\frac{2M}{r}\Big)-\frac{\beta m_a}{M_Pr}e^{-m_ar}-\frac{\beta}{r^2M_P}e^{-m_a r},
\label{eq:29}
\end{equation} 
where $\dot{r}=\frac{dr}{d\lambda}=\frac{E}{\Big(1-\frac{2M}{r}\Big)}\frac{dr}{dt}$. Therefore, we can write Eq. \ref{eq:29} as
\begin{equation}
\frac{E^2}{2}=\frac{E^2}{2\Big(1-\frac{2M}{r}\Big)^2}\Big(\frac{dr}{dt}\Big)^2+\frac{L^2}{2r^2}\Big(1-\frac{2M}{r}\Big)-\frac{\beta m_a}{M_P r}e^{-m_a r}-\frac{\beta}{r^2M_P}e^{-m_ar}.
\label{eq:30}
\end{equation}
Let, $r=r_0$ is the closest approach of light where $\frac{dr}{dt}=0$. Put, $r=r_0$ and $\frac{dr}{dt}=0$ in Eq. \ref{eq:30} we obtain
\begin{equation}
\frac{L^2}{E^2}=\frac{r^2_0}{\Big(1-\frac{2M}{r_0}\Big)}\Big[1+\frac{2\beta}{M_P r_0E^2}\Big(m_a+\frac{1}{r_0}\Big)e^{-m_a r_0}\Big].
\label{eq:31}
\end{equation}
In absence of monopole-dipole potential, Eq. \ref{eq:31} becomes $\frac{L^2}{E^2}=\frac{r^2_0}{\Big(1-\frac{2M}{r_0}\Big)}$. Using Eq. \ref{eq:30} and Eq. \ref{eq:31}, we can write the time taken by the light to reach from $r_0$ to $r$ as 
\begin{equation}
t=\int^r_{r_0}\frac{dt}{dr}dr=\int^r_{r_0}dr\frac{1}{\Big(1-\frac{2M}{r}\Big)}\Big[1-\frac{r^2_0}{r^2}\frac{\Big(1-\frac{2M}{r}\Big)}{\Big(1-\frac{2M}{r_0}\Big)}(1+\eta)+\frac{2\beta}{M_PrE^2}\Big(m_a+\frac{1}{r}\Big)e^{-m_ar}\Big]^{-\frac{1}{2}},
\label{eq:32}
\end{equation}
where $\eta=\frac{2\beta}{M_P r_0E^2}\Big(m_a+\frac{1}{r_0}\Big)e^{-m_ar_0}$. The solution of Eq. \ref{eq:32} in $r\gg r_0$ limit is
\begin{equation}
t_1=\sqrt{r^2-r^2_0}+2M\ln\Big(\frac{2r}{r_0}\Big)+M-\frac{\beta}{M_PE^2}\Big(\frac{M}{r^2_0}+\frac{1}{r_0}\Big)+\frac{\eta r_0}{2}\Big(1+\frac{2M}{r_0}\Big).
\label{eq:33}
\end{equation}
If $r_e$ denotes the distance between Sun and Earth and $r_v$ denotes the distance between Sun and Venus then the total time required for the signal to go from Earth to Venus and returns to Earth is
\begin{equation}
\begin{split}
T_1=2t_1=2\Big[\sqrt{r^2_e-r^2_0}+\sqrt{r^2_v-r^2_0}+2M\ln\Big(\frac{2r_e}{r_0}\Big)+2M\ln\Big(\frac{2r_v}{r_0}\Big)+2M-\frac{2\beta M}{M_PE^2r^2_0}-\\
\frac{2\beta}{M_PE^2r_0}+\eta r_0\Big(1+\frac{2M}{r_0}\Big)\Big].
\end{split}
\label{eq:34}
\end{equation} 
If there is no massive gravitating object between Earth and Venus, then the total time required for the pulse to go from Earth to Venus and returns to Earth is
\begin{equation}
T_2=2\Big[\sqrt{r^2_e-r^2_0}+\sqrt{r^2_v-r^2_0}-\frac{2\beta}{M_PE^2r_0}+\eta r_0\Big].
\label{eq:35}
\end{equation}
Hence, the excess time due to GR correction and monopole-dipole potential is $\Delta T=T_1-T_2$ and we can write
\begin{equation}
\begin{split}
\Delta T=4M\Big[1+\ln\Big(\frac{4r_er_v}{r^2_0}\Big)\Big]-\frac{4M}{M_PE^2r^2_0}\Big(\frac{g_Sg_PN_1N_2}{4\pi m_e}\Big)+\frac{8M}{M_Pr_0E^2}\Big(m_a+\frac{1}{r_0}\Big)e^{-m_ar_0}\Big(\frac{g_Sg_PN_1N_2}{4\pi m_e}\Big),
\end{split}
\label{eq:36}
\end{equation}
where we put the expressions of $\beta$ and $\eta$. If there is no monopole-dipole potential then $g_Sg_P\rightarrow 0$ and we get back the standard GR contribution in Shapiro time delay as 
\begin{equation}
\Delta T_{\rm{GR}}=4M\Big[1+\ln\Big(\frac{4r_er_v}{r^2_0}\Big)\Big].
\label{eq:37}
\end{equation}
Using the Earth-Sun distance $r_e=D=1.46\times 10^{11}~\rm{m}=7.37\times 10^{26}~\rm{GeV^{-1}}$, the Venus-Earth distance $r_v=1.08\times 10^{11}~\rm{m}=5.47\times 10^{26}~\rm{GeV^{-1}}$, and the solar radius $r_0=R_\odot=6.96\times 10^8~\rm{m}=3.51\times 10^{24}~\rm{GeV^{-1}}$, we obtain the GR contribution in Shapiro time delay as $2\times 10^{-4}~\rm{s}$. Thus the contribution of monopole-dipole potential in Shapiro time delay is
\begin{equation}
\begin{split}
\Delta T_{\rm{monopole-dipole}}=\frac{8M}{M_Pr_0E^2}\Big(m_a+\frac{1}{r_0}\Big)e^{-m_ar_0}\Big(\frac{g_Sg_PN_1N_2}{4\pi m_e}\Big)-\frac{4M}{M_PE^2r^2_0}\Big(\frac{g_Sg_PN_1N_2}{4\pi m_e}\Big)+\\
\mathcal{O}\Big((g_Sg_P)^2, m^2_a, M^2\Big).
\end{split}
\end{equation}
\bibliographystyle{utphys}
\bibliography{md}
\end{document}